\documentclass[]{article}
\usepackage{authblk}

\usepackage{graphics}
\usepackage{graphicx}
\usepackage{verbatim}
\usepackage{acronym}
\usepackage{color}
\usepackage{comment}
\definecolor{mygreen}{rgb}{0,0.6,0}
\definecolor{mygray}{rgb}{0.5,0.5,0.5}
\definecolor{mymauve}{rgb}{0.58,0,0.82}
\usepackage{subcaption}
\usepackage[T1]{fontenc}
\usepackage[english]{babel}
\usepackage{amsmath,amsfonts,amsthm}

\usepackage{amssymb}
\usepackage{algorithmicx}
\usepackage{algorithm}
\usepackage[noend]{algpseudocode}
\usepackage{csquotes}
\usepackage{url}
\usepackage{paralist}
\usepackage{verbatimbox}
\usepackage{verbdef}

\usepackage{pgfplots}
\usepackage{tikz}
\usetikzlibrary{shapes,arrows,automata,positioning}
\pgfmathdeclarefunction{gauss}{2}{%
  \pgfmathparse{ (\yoffset+1/(sqrt(2*pi*#2))*exp(-((x-#1)^2)/(2*#2)) )}%
}

\makeatletter \newcommand{\pgfplotsdrawaxis}{\pgfplots@draw@axis} \makeatother

\pgfplotsset{
tick label style = {font=\sansmath\sffamily},
every axis label/.append style={font=\sffamily\footnotesize},
}

\pgfplotsset{axis line on top/.style={
  axis line style=transparent,
  ticklabel style=transparent,
  tick style=transparent,
  axis on top=false,
  after end axis/.append code={
    \pgfplotsset{axis line style=transparent,
      ticklabel style=opaque,
      tick style=opaque,
      grid=none}
    \pgfplotsdrawaxis}
  }
}

\tikzset{>=latex}

\newcommand{\meanI}{62.5}
\newcommand{\varI}{80}

\newcommand{\labelheight}{0.015}

\pgfmathsetmacro{\meanIheight}{1/(sqrt(2*pi*\varI))}%
\pgfmathsetmacro{\plotheight}{max(\meanIheight+.01, \labelheight+0.01)}%

\newcommand{\yoffset}{0.0075}

\verbdef{\vtext}{verb text}

\newacro{MVDA}{MultiVariate Data Analysis}
\newacro{ANOVA}{ANalysis Of VAriance}
\newacro{ATP}{Adenosine TriPhosphate}
\newacro{BRCA}{BReast CAncer Gene}
\newacro{CLT}{Central Limit Theorem}
\newacro{DM2}{Diabetes Mellitus 2}
\newacro{DNA}{DeoxyriboNucleic Acid}
\newacro{ERPR}{Estrogen Receptor Progresterone Receptor}
\newacro{FDR}{False Discovery Rate}
\newacro{FOX}{Forkhead box}
\newacro{FPKM}{Fragments Per Kilobase of transcript per Million mapped reads}
\newacro{FR}{Fractional Rank}
\newacro{GEO}{Gene Expression Omnibus}
\newacro{GTP}{Guanosine TriPhosphate}
\newacro{GWAS}{Genome Wide Association Study}
\newacro{HTS}{High Throughput Sequencing}
\newacro{mRNA}{messenger Ribonucleic Acid}
\newacro{MS}{Multiple Sclerosis}
\newacro{NGT}{Normal Glucose Tolerance}
\newacro{OPLS}{Orthogonal Projection to Latent Structures}
\newacro{PLS}{Partial Least Squares}
\newacro{EPLS}{Exploratory Projection to Latent Structures}
\newacro{RMSE}{Root Mean Squared Error}
\newacro{PCA}{Principle Component Analysis}
\newacro{RNA}{Ribonucleic Acid}
\newacro{TCGA}{The Cancer Genome Atlas}
\newacro{T2D}{Type 2 Diabetes}

\usepackage{listings}
\usepackage{hyperref}

\newcommand{\opls}{Exploratory Projection to Latent Structure Models for use in Transcriptomic Analysis}
\newcommand{\norm}[1]{||#1||}
\newcommand{\mat}[1]{\mathbf{#1}}
\newcommand{\gene}[1]{\texttt{#1}}
\newcommand{\furl}[2]{\href{#2}{#1}\footnote{\url{#2}}}
\renewcommand{\vec}[1]{\mathbf{#1}}

\providecommand{\keywords}[1]
{
  \small
  \textbf{\textit{Keywords---}} #1
}

\begin{document}

\title{\opls}

\author[$\dagger$,$\ddagger$]{Edward Tj{\"{o}}rnhammar}
\author[$\star$]{Richard Tj{\"{o}}rnhammar}

\affil[$\dagger$]{KTH Royal Institute of Technology, SE-100 44 Stockholm, Sweden}
\affil[$\ddagger$]{FOI Swedish Defence Research Agency, SE-164 90 Stockholm, Sweden}
\affil[$\star$]{KI Karolinska Institutet, SE-171 71 Stockholm, Sweden}

\maketitle

\begin{abstract}

In this paper, we ask if it is possible to increase the interpretability in
multivariate analysis by aligning and projecting covariates onto comparative
subspaces. We demonstrate our method as well as the interpretative power of
\ac{PLS} decomposed models and how robust interpretability can lead to
quantitative insights.

We discuss the statistical properties of the \ac{PLS} weights, $p$-values
associated with specific axes, as well as their alignment properties.

The applicability of this approach within life science is also demonstrated by
applying it to three use cases of publically available datasets.

Further we present hierarchical pathway enrichment results stemming from aligned
$p$-values, which are compared with results derived from enrichment analysis, as
an external validation of our method.

We find that the method can uncover known results from genomics for all of the
studied use cases, i.e. microarray data from multiple sclerosis and diabetes
patients as well as RNA sequencing data from breast cancer patients.

\end{abstract}


\section{Introduction}

Many fields within Life Science, such as transcriptomics (the study of
transcripts), proteomics (the study of proteins), or metabolomics (the study of
metabolites) characteristically rely upon datasets which has ``few samples'' but
``many features''. One is furthermore typically interested in discovering which
collection of analytes are modulating specific attributes.

\ac{OPLS} models~\cite{wold_multivariate_1984,wegelin_survey_2000} have a long
history of usage in
metabolomics~\cite{rothenberg_metabolome_2019,eriksson_multi-_2013}, as well as
other fields of research. The ability to decompose and align the feature space
while communicating the effect of the decomposition on sample space is a
well-known property of \ac{OPLS} models. The \ac{OPLS} methods have long been
employed for characterisation of data features, but the interpretation often
stops once the sample and feature groupings have been identified.

Both \ac{PCA}~\cite{abdi_principal_2010} and \ac{OPLS} create new predictor
variables often termed components. The components created by \ac{PCA} maximize
the variance in each without regard for the response. In \ac{PCA}, we get a
large number of components that aim to describe orthogonal variance. This affine
transform can be further exploited to evaluate the statistical significance
of weight relations. Similar to \ac{PCA}, the \ac{OPLS} decomposes an input
matrix into unique and complementary matrices.
However, unlike \ac{PCA}, which ultimately relies on factorising a sample
centred expression matrix along the components of maximal orthogonal variance,
the \ac{OPLS} method instead decomposes an input matrix by maximizing the mutual
covariance between the feature space projection against the sample space
response projection on the \ac{OPLS} predictor plane.

The subspace plane could be any dimensionality and in \ac{PLS} regression the
NIPALS algorithm ensures that the covariance between the score and weight vectors
are aligned.~\cite{rosipal_overview_2006} (second equation). Because the weights
are commonly computed via an iterative procedure it does not necessarily
converge to a good solution, it however often does. Once the solution has been
obtained then the variance in the dataset is often better described by fewer
components than in \ac{PCA}. It is not strange because the goal is to model the
response by directly decomposing variance onto components describing the
response. For this reason, a \ac{OPLS} algorithm is similar to a two-way linear
regression \ac{ANOVA}~\cite{fisher_xv.correlation_1919}. In both methods a two-dimensional
tensor is constructed to describe the response and as such both
methods are prone to similar issues and produce similar statistical insights. We
will see that, as opposed to a two-way \ac{ANOVA} model, the \ac{OPLS} model also
divulge information about the directionality of the variation along an axis.

In~\cite{tapp_notes_2009} the authors make a point that \ac{OPLS} regression
coefficients used in ranked analysis do not take advantage of the multivariate
nature of \ac{OPLS}. This holds even for our work, we project the decomposition
down onto a univariate vector to increase the robustness of our model
interpretations.

We define a pathway as a knowledge-based group descriptor of the
transcripts. It should not be confused with the path diagram of the latent
variables~\cite{wegelin_survey_2000}. Connecting the weights with axis
$p$-values and calculating pathway enrichment becomes an important sanity check
for this method. There are several knowledge-based pathway
databases~\cite{szklarczyk_string_2019,kanehisa_kegg_2000,jassal_reactome_2019,mi_panther_2018}.
These are curated with regards to known activity of transcripts and as such are
not deduced from knowledge inherent in the data. In theory a new and properly
well carried out experiment should be unknown to the knowledge database curator
facilitating the group definitions. As such, if the alignment of weights to a
descriptor axis is sensible then, the pathway enrichments should convey sensible
enrichment for the groups that are under study.

For our purposes; Given the feature and sample space description matrices
$(\mat{X}, \mat{Y})$ are described by their corresponding weights
$(\mat{W},\mat{C})$ and scores
$(\mat{T}=\mat{X}\mat{W},\mat{U}=\mat{Y}\mat{C})$. If the \ac{PLS} algorithm
has converged on a solution, by maximizing covariance between $\mat{T}$ and
$\mat{U}$, then $\mat{w}$ and $\mat{c}$ will be aligned on the projected
subspace~\cite{rosipal_overview_2006}.

In this work we propose a method for decision research support in noisy domains.
We believe it to be able to find plausible research targets by an exploration of
the feature space encoding in high entropy domains through comparative plotting
of cosine alignment on \ac{PLS} decomposition. This method is similar to
\ac{OPLS}~\cite{trygg_orthogonal_2002}, but it does not perform discriminant
analysis, such as in \ac{OPLS}-DA, which facilitates the analysis of categorical
data.  Using this approach, with encoded information and utilizing the $\mat{C}$
space as an attribute compass in the feature weight plane one is able to compute
$p$-values for sample attributes in a \ac{PLS} context.

As such we demonstrate that the alignment \ac{PLS} decomposition can be directly
utilized for $p$-value calculations. We also demonstrate that large values along
the attribute axis corresponds to large fold change contribution between groups
along said axis.

We believe the main contributions of the paper to be:
\begin{itemize}
    \item An exploratory technique combining alignment and angular
          representation of samples with regards to some distance measure. The
          exploratory \ac{PLS} method creates a robust multidimensional fit
          whereby signal artifacts not aligned with the encoding obtain small
          weights in their component.
    \item An evaluation of the method with regards to comparable analyses, i.e.
          T-test, ANOVA and PCA, as well as independent boxplot validation.
    \item A demonstration of said method with regards to already known results.
    \item A multivariate regression method implemented in python.
\end{itemize}

The remainder of the paper is structured as follows. Section~\ref{sec:method}
motivates and describes the statistical viewpoint and exemplifies analysis
usage. A number of illustrating analysis examples related to the previously
described method process are then presented in Section~\ref{sec:results} in
order to exemplify and validate the proposed solution. For validation purposes a
comparison to t-testing is performed. Finally, Sections~\ref{sec:discuss}
and~\ref{sec:conclusions} provide an assessment analysis and conclude from the
undertaken experiments.

\section{Method} \label{sec:method}

\begin{figure}[ht]
  \begin{center}

    \tikzstyle{plotA}=[ultra thick,red!90!black]

    \begin{tikzpicture}[>=stealth',
      el/.style = {inner sep=8pt, align=left, sloped},
      every label/.append style = {font=\tiny}]
      \usetikzlibrary{fit}
      \tikzstyle{decision} = [diamond, draw, fill=blue!20,
        text width=5em, text badly centered, node distance=3.5cm, inner sep=0pt]

      \tikzstyle{main}=[circle, minimum size = 10mm, thick, draw =black!80, node distance = 16mm]
      \tikzstyle{part}=[ellipse, minimum size = 10mm, thick, draw =black!80, node distance = 16mm]
      \tikzstyle{indata}=[rectangle, minimum size = 5mm, thick, draw =black!80, fill = black!10, node distance = 16mm]
      \tikzstyle{connect}=[-latex]
      \tikzstyle{dashing}=[dashed]
      \tikzstyle{selector}=[-latex, -|, snake=snake,segment amplitude=.4mm,segment length=2mm,line after snake=1mm, thick]
      \tikzstyle{shortconnect}=[-latex, thin]
      \tikzstyle{box}=[rectangle, draw=black!100]
      \tikzstyle{switch}=[circle, minimum size = 1mm, fill = black!100, draw=black!100]

      \node[indata] (inpform) [] {Formula: $f$};
      \node[indata] (inpjournal) [left of=inpform, xshift=-4mm, label=\text{coarse grained data}] {Journal: $J$};
      \node[part] (encode) [below of=inpform,yshift=-10mm, label=below:Encoding] {$f(J) \rightarrow E$};
      \node[part] (pls) [right of=encode,xshift=20mm] {PLS$(E,A)$};
      \node[indata] (analytes) [above of=pls,xshift=20mm, label=\text{fine grained data}] {Analytes: $A$};

      \node[part] (alignment) [right of=pls,yshift=-30mm, label=below:] {Alignment};
      \node[part] (projection) [right of=alignment,xshift=20mm, label=below:] {Projection};

      \begin{axis}[
          at={(0.475\linewidth,-0.30\linewidth)},
          width=6.0cm, 
          height=3.0cm, 
          grid = major,
          grid style=\empty, 
          xmin= 0,     
          xmax= 125,   
          ymin= 0,     
          ymax= \plotheight, 
          legend style={
              font=\empty, 
              at={(0.5,-0.18)},
              anchor=north,
              legend cell align=left,
              legend columns=-1,
              column sep=0.5cm
          },
          tick align=inside,
          scaled y ticks = false,
          ytick=\empty,
          axis lines = middle,
          axis line style = thick,
          axis line on top,
          axis line style = {thick,shorten <=-0.5\pgflinewidth},
          xlabel=attribute axis,
          x label style={at={(axis description cs:0.52,0)},
                         anchor=north,
                         font=\boldmath},
          xtick=\empty,
          ylabel=$ $,
          y label style={at={(axis description cs:0,0.5)},
                         anchor=south,
                         rotate=90,
                         font=\boldmath\Large},
        ]
        \coordinate (axisOrigin) at (axis cs:0,0);
        \coordinate (meanA) at (axis cs:\meanI,\meanIheight);
        \draw[black,<->](axis cs:{\meanI-4*sqrt(\varI)}, 0.5*\yoffset) -- (axis cs:{\meanI+4*sqrt(\varI)}, 0.5*\yoffset);
        \addplot [domain=\meanI-4*sqrt(\varI):\meanI+4*sqrt(\varI),samples=100,plotA] {gauss(\meanI,\varI)};
        \node[plotA,below] (gaussg) at (axis cs:\meanI,\labelheight+\yoffset){$p$-values};
        \node[box,minimum size=25mm,draw=white!0] (gaussian) at (axis cs:\meanI,\labelheight+\yoffset){};
      \end{axis}

      \node[decision] (kmeans) [left of=alignment,xshift=-10mm,] {KNN $K=1$ where $\vec{c}$ are centroids};

      \path (inpform) edge [connect] (encode)
      (inpjournal) edge [connect] (encode)
      (encode) edge [connect] node[el,below] {%
        $\Big[\tiny\begin{smallmatrix}%
          0 & 1 & 1 & 0 \\ %
          1 & 0 & 0 & 1 \\
          0.5 & 5 & 4.1 & 0.1 %
        \end{smallmatrix}\normalsize\Big]$
        } (pls)
      (analytes) edge [connect] (pls)
      (pls) edge [connect] node[el,below] {\tiny $(\mat{W},\mat{T},\mat{C},\mat{U})$ \normalsize} (alignment)
      (kmeans) edge [dashing] (alignment)
      (alignment) edge [connect] (projection)
      (projection) edge [dashing] (gaussian);
      \node[rectangle, inner sep=4.4mm,draw=black!100, fit= (alignment) (projection),label=below right:EPLS:Projection properties] {};
      \node[rectangle, inner sep=4.4mm,draw=black!100, fit= (encode) (pls),label=below right:EPLS:Regression] {};
    \end{tikzpicture}
    \caption{Method simplified workflow}
    \label{fig:epls-method-graph}
  \end{center}
\end{figure}
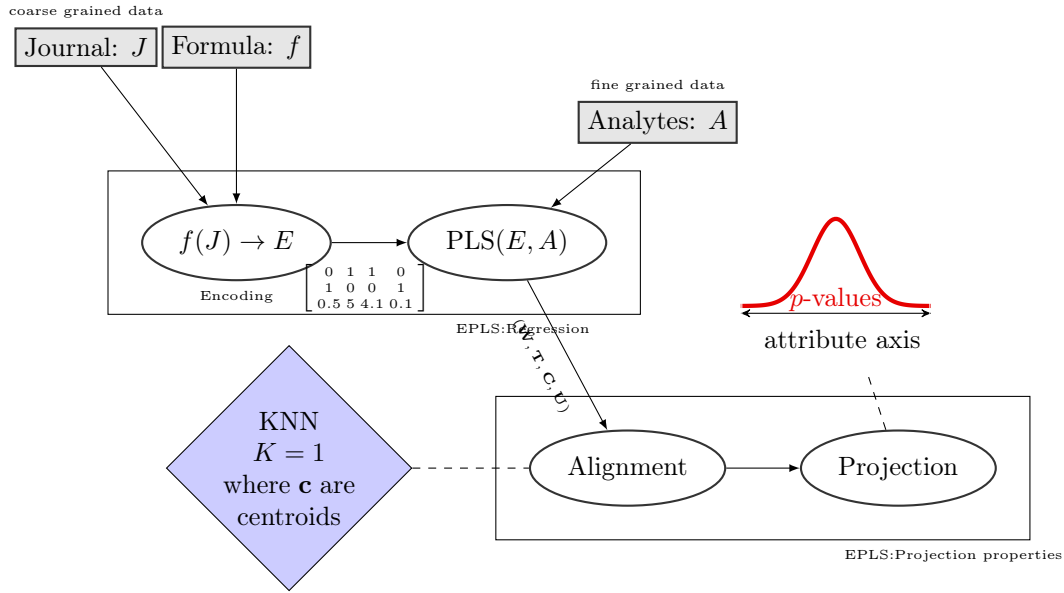

For our adaptation of the \ac{PLS} method, \ac{EPLS}, we employ the same
nomenclature as in the work by~\cite{rosipal_overview_2006}. We present an
overview of the method in Figure~\ref{fig:epls-method-graph}.

It should be noted that the journal $J$ and ``analytes'' $A$ are respectively
previously denoted $X$ and $Y$ in regular \ac{PLS}. The journal $J$ is simply
named in this way since it is a record of the different patients and their
biometrics.

We begin by specifying the encoding formula for the problem at hand. A formula
is on the form:

\[
  \text{y} \sim C(x_1):C(x_2) + C(x_3) + x_4
\]

Meaning that $y$ is inspected by encoding an $X$ matrix constituting of the
values of $x_4$ from $J$, the binary class expansion of $x_3$ from $J$ and a
binary class expansion of the binomial pair combinations of $x_1$ and $x_2$ from
$J$. With a binary class expansion we mean that $C$ encodes patient $p$
``diabetic'' biometric marker status as 1 or 0. As a concrete example we might
have the patient journal as in Table~\ref{tab:journal}. Which according formula,
assuming $x_1 = \text{Diabetic}, x_2 = \text{Smoker} \ldots$,  would encode into
Table~\ref{tab:formulaencodedmx}

\begin{table}[t]
\begin{tabular}{lllll}
Diabetic    & F   & T$_2$ & T$_1$   & F   \\
Smoker      & S   & S     & F       & F   \\
Unemployed  & F   & S     & F       & F   \\
Blood-sugar & 0.5 & 5     & 4.1     & 0.1
\end{tabular}
\caption{Example patient $J$ournal}
\label{tab:journal}
\end{table}

\begin{table}[b]
\begin{tabular}{lllll}
Diabetic.T$_1$-Smoker.F   & 0   & 0 & 1   & 0   \\
Diabetic.T$_2$-Smoker.S   & 0   & 1 & 0   & 0   \\
Diabetic.F-Smoker.F       & 0   & 0 & 0   & 1   \\
Diabetic.F-Smoker.S       & 1   & 0 & 0   & 0   \\
Unemployed.F              & 1   & 0 & 1   & 1   \\
Unemployed.S              & 0   & 1 & 0   & 0   \\
Blood-sugar               & 0.5 & 5 & 4.1 & 0.1
\end{tabular}
\caption{Coded patient Table~\ref{tab:journal} $J$ournal}
\label{tab:formulaencodedmx}
\end{table}

As one can discern from Table~\ref{tab:formulaencodedmx} that the attributes are
exploded into their value form and multi-class attributes, i.e. Diabetic here,
compounded with the other attribute's value form.

Given the subspace vectors $\mat{W} \wedge \mat{T}$ we assume that any symmetry
operations on $\mat{W}$ will commute to $\mat{T}$. Habitually for our numerical
experiments, we treat a set of gene expressions as the feature matrix $\mat{X}$
and the patient journal, with coarse-grained end states, as the sample response
matrix $\mat{Y}$. As such $\mat{X}$ either represent a microarray- or a
\ac{FPKM} \ac{RNA}-sequence matrix.

The formulation of groups is done by retrieving journal value entries for a
group present in the statistical formula enclosed with parentheses and a leading
`$C$'. Example:
\[
  \text{Transcript} \sim C(\text{\gene{Group}})
\]
would retrieve the \texttt{Group} journal value entries.

To elaborate, the formula describes the

Since \ac{PLS}2 maximizes the pairwise covariance score, feature weights and
sample weights are ``co-linearized'' in a similar way. This lets us ascribe
features to sample weights, one way is to employ tessellation. Each transcript
is simply associated with the closest ``formulae encoded axis''. The formulae
encoded axes are defined by the PLS regression weights, using the above
notation, of the $Y$ matrix. We will refer to this target space as the ``sample
axis''. We haven't proved this property since it is assured by the PLS
regression work by~\cite{rosipal_overview_2006}.

For proximity we can either use absolute coordinate proximity on the weight
projection plane or the angular proximity between the transcript and attribute axis
weight. The formula for calculating the coordinate proximity follows $1-NN$
where the centroids are taken from the \ac{PLS} $\mat{Y}$ estimation $\mat{C}$.
The formula for calculating the angular distance is given by the cosine
similarity:

\[
d = 1 - \cos(\alpha) = 1 - \frac{\vec{a} \cdot \vec{b}}{\norm{\vec{a}}\cdot\norm{\vec{b}}}
\]

Here $\vec{a}$, a vector, designates a transcript weight position and $\vec{b}$
a sample score weight position, also a vector. The angle $\alpha$ is the angle
between $\vec{a}$ and $\vec{b}$. In summary, we perform the steps:

\begin{enumerate}
  \item Use a statistical test formula that defines the group variations to
        test, i.e. $f$.
  \item Device an encoding data frame $\mat{E}$ from journal group descriptor $\mat{J}$
        instances, i.e. $f(\mat{J}) \rightarrow \mat{E}$.
  \item Utilize PLS2 to find the solution to the formula using the expression
        matrix and patient (sample response) journal, i.e.
        $PLS(\mat{E},\mat{A}) \rightarrow (\mat{W},\mat{T},\mat{C},\mat{U})$
  \item Categorical membership is determined through either tessellation
        (euclidean distance) or angle, cosine distance (1-NN) to a centroid, the
        centroids are given by $\vec{c}$, which denotes a vector belonging to $\mat{C}$.
  \item Calculate characteristic length scales of the prediction plane by
        finding the maximum weights for normalization purposes across the PLS components.
  \item Calculate the projected density of the weights onto the group axes
        as encoded by the statistical test formula, i.e.
        $ \forall \vec{c} \in \mat{C} \wedge \vec{w} \in \mat{W}$ we calculate
        the projections $ r = \frac{\vec{c} \cdot \vec{w}}{\left\Vert\vec{c}\right\Vert} $
  \item Calculate $p$-values for the projections beloning to a $\vec{c}$ by assuming that the weights
        are distributed normally, using the complementary cumulative distribution function
        for the observed projection,
        $Q(r)=1-\frac{1}{2}\left[1+erf(\frac{r-\mu}{\sigma \sqrt2}) \right]$.
  \item Calculate the ranking of the weights on the projected axis, mapped in
        ascending order from zero to one, i.e.
        $\#\{i: |r_i \leqslant r_j, \forall (i,j) \in \vec{c}\}$
\end{enumerate}

This can be relied upon to visualize how features associate with samples.

Since ReactomePA~\cite{jassal_reactome_2019} knowledge base is hierarchical, we
also employed a hierarchical enrichment and correction scheme similar to the
elim algorithm as suggested by~\cite{alexa_improved_2006}.

\subsection{Implementation} \label{sec:impl}

We have chosen to employ the \ac{PLS}2 algorithm as implemented in the
publicly available python~\cite{pedregosa_scikit-learn_2011} package and it
also exists as an R~\cite{geladi_partial_1986,hoskuldsson_pls_1988} package.

All categoricals present in the patient journal are translated into a binary
encoding matrix. Thus all and any unique descriptors included in the group under
study gets translated into a unique axis. All formula entries that are not
categoricals are treated as real variables. The exact routine execution,
visualization and calibration is covered in the experiment
code~\cite{tjornhammar_experiments_2020}.

The quantitative analysis relies on projecting the feature weight distribution
onto an axis and studying the point density distribution on that axis. The $p$
values are calculated from the cumulative error function for the density along
the projection axis, see supplementary
code~\cite{tjornhammar_impetuous-gfa_2019}. The two-way \ac{ANOVA} comparison is
performed using the statsmodels~\cite{seabold_statsmodels_2010} SciPy package.

\section{Results} \label{sec:results}

According to the method described in Section~\ref{sec:method} we conducted three
experiments:

\begin{itemize}
    \item \ac{MS} microarray dataset analysis, testing sample variation to group
    interaction. Also presenting the alignment properties of angular \ac{EPLS} model
    interpretation.
    \item \ac{TCGA}-\ac{BRCA} two-way \ac{ANOVA}, deducing the interaction model
    from the \ac{EPLS} plot. Also comparing the analytical strength to type-2
    \ac{ANOVA}.
    \item Diabetes MICROARRAY dataset compounded analysis with clustering
     enrichment for genomic impact in \ac{T2D}.
\end{itemize}

The datasets for the analysis, as well as the group definition files, can
all be found online~\cite{tjornhammar_impetuous-gfa_2019}. Details of the
processing of the \ac{MS}, as well as \ac{TCGA} dataset, is also deposited
online~\cite{tjornhammar_ms_2019}.

\subsection{Multiple Sclerosis and the alignment properties of \ac{EPLS} models} \label{sec:results:ms}

The Multiple Sclerosis microarray data was obtained via the \ac{GEO} accession
number for the microarray dataset
\furl{GSE21942}{http://www.ncbi.nlm.nih.gov/geo/}. 15 `\texttt{Healthy}'
controls and 14 `\texttt{MS}' patients were processed and used in the
analysis~\cite{kemppinen_systematic_2011}. The \ac{EPLS} model is conducting the
test whether a transcript sample variation is well described by the
between-group variation. We choose to express this as a statistical formula on the form
\[
  \text{Transcript} \sim C(\text{\gene{Status}})
\]
Here the \texttt{Status} label can take on the values
\texttt{Healthy} or \texttt{MS}.

\begin{figure*}[t!]
  \centering
  \begin{subfigure}[t]{0.45\textwidth}
    \centering
    \includegraphics[scale=0.3]{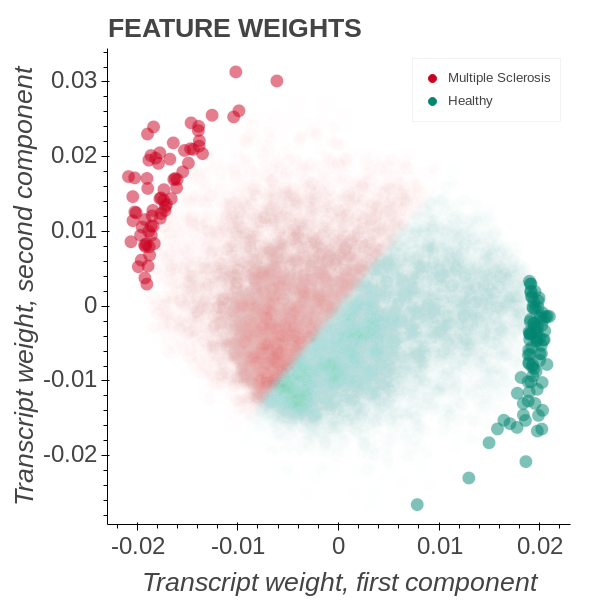}
    \caption{The EPLS transcript weight plane with weights coloured by alignment to \ac{MS} state}
    \label{fig:opls-ms-exploration-trans}
  \end{subfigure}
  ~
  \begin{subfigure}[t]{0.45\textwidth}
    \centering
    \includegraphics[scale=0.3]{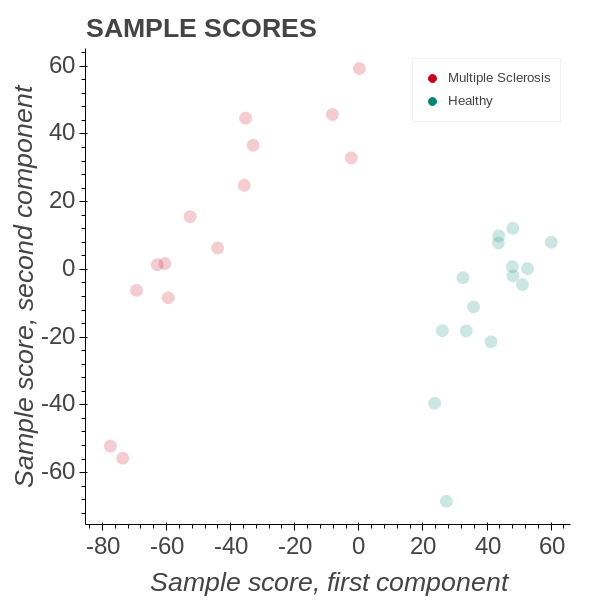}
    \caption{The EPLS sample scores with samples coloured by \ac{MS} state}
    \label{fig:opls-ms-exploration-sample}
  \end{subfigure}
  \caption{EPLS model of the \ac{MS} state.}
  \label{fig:opls-ms-exploration}
\end{figure*}

In Figure~\ref{fig:opls-ms-exploration-trans} we can see how the transcripts
become aligned to the disease state. A large number of transcripts are
positioned close to origo and those that are farther away than 99\% of all
transcripts are highlighted with a higher opacity. From
Figure~\ref{fig:opls-ms-exploration-sample}, it is clear that the \ac{EPLS}
model separates the cohort well. By comparing the left and right graph we see
that the transcript weights are aligned so that we only have significant weights
in the same quadrants as samples. The $\mat{Y}$ matrix, describing the
\texttt{MS} and \texttt{Healthy} subjects sample score weights are also in the
same quadrants. We can calculate the angular proximity between the sample score
weights and the transcripts feature weights and assign what transcripts are
closest to a particular score weight. This allows us to colour the transcripts
according to the descriptor that best captures their variation in the data. We
can also define new vectors corresponding to different directions on the
transcript weight plane.

\begin{figure*}[t!]
  \centering
  \begin{subfigure}[t]{0.45\textwidth}
    \centering
    \includegraphics[scale=0.3]{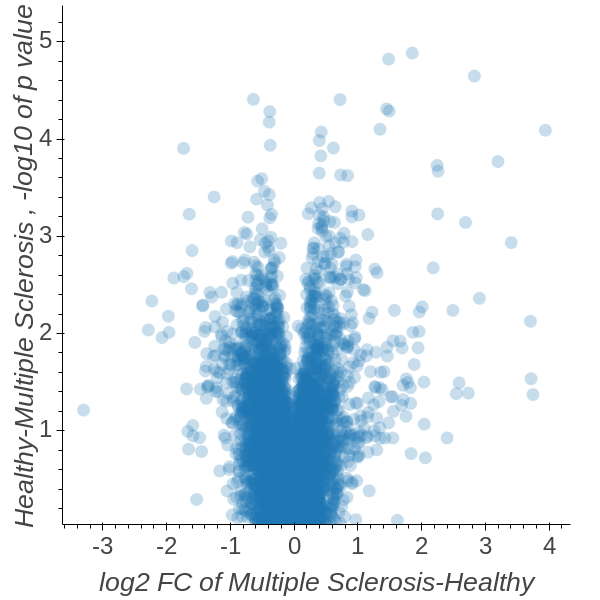}
    \caption{The gene transcripts associated \ac{EPLS} $p$ values against $log_2$ fold changes.}
    \label{fig:opls-hms-trans}
  \end{subfigure}
  ~
  \begin{subfigure}[t]{0.45\textwidth}
    \centering
    \includegraphics[scale=0.5]{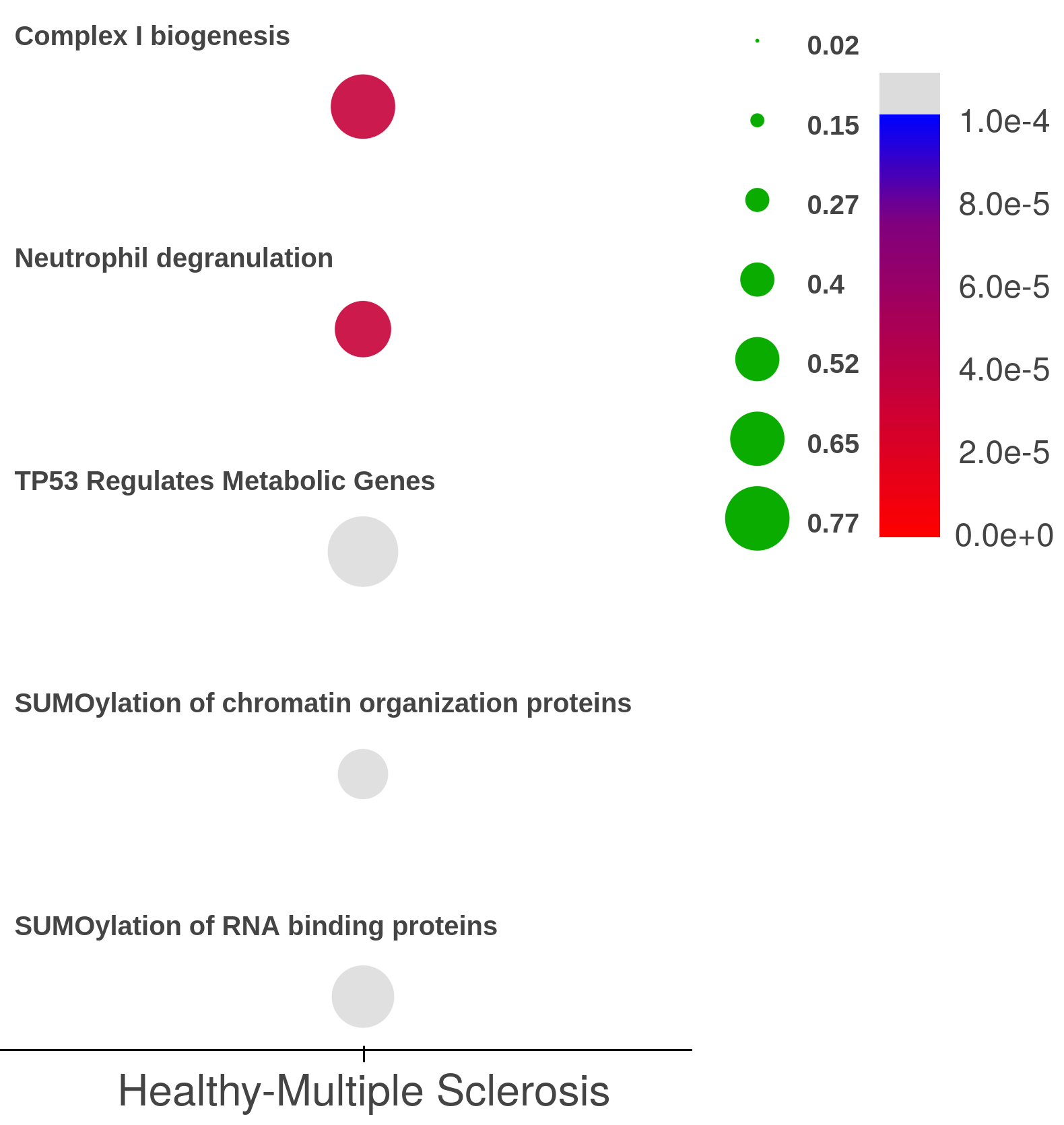}
    \caption{Hierarchically corrected pathway enrichment. }
    \label{fig:opls-hms-hier}
  \end{subfigure}
  \caption{Sizes corresponds to gene ratios and the significance levels depicted
    are raw $p$ values.}
  \label{fig:opls-hms-exploration}
\end{figure*}

In Figure~\ref{fig:opls-hms-trans}, we depict the negative logarithm of the
\ac{EPLS} $p$-values on the $y$-axis and fold change values on the $x$-axis. In
this case, the \ac{EPLS} model describes the separation between \texttt{Healthy}
and \texttt{MS} patients and the $p$ values are thus analogue to $t$-test
comparison based $p$ values resulting in a volcano plot appearance. The axis
describing the Healthy-Multiple Sclerosis state variation $p$ values can then be
employed to calculate pathway enrichments. In Figure~\ref{fig:opls-hms-hier},
the pathway enrichment has been calculated using a Fisher exact test together
with the Reactome pathway database. For each pathway, significant genes are
deduced based on a $p$-value cutoff equal to $0.05$ and a contingency table is
produced. The resulting contingency table is then evaluated using a two-sided
Fisher exact test. We choose a two-sided test because we employ a variational
model. As such the enrichment does not correspond to an up or down-regulated
accumulation in the group but rather correspond to if the group is important in
describing the \ac{EPLS} axis.

\begin{figure*}[t!]
  \centering
  \begin{subfigure}[t]{0.45\textwidth}
    \centering
    \includegraphics[scale=0.3]{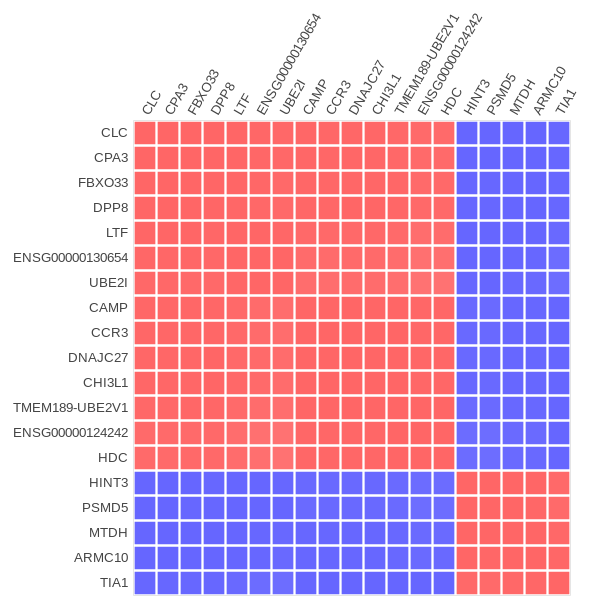}
    \caption{The gene transcripts association map. Colors correspond to the
      cosine of the angle between the feature weights.}
    \label{fig:gene-cosine-assoc}
  \end{subfigure}
  ~
  \begin{subfigure}[t]{0.45\textwidth}
    \centering
    \includegraphics[scale=0.5]{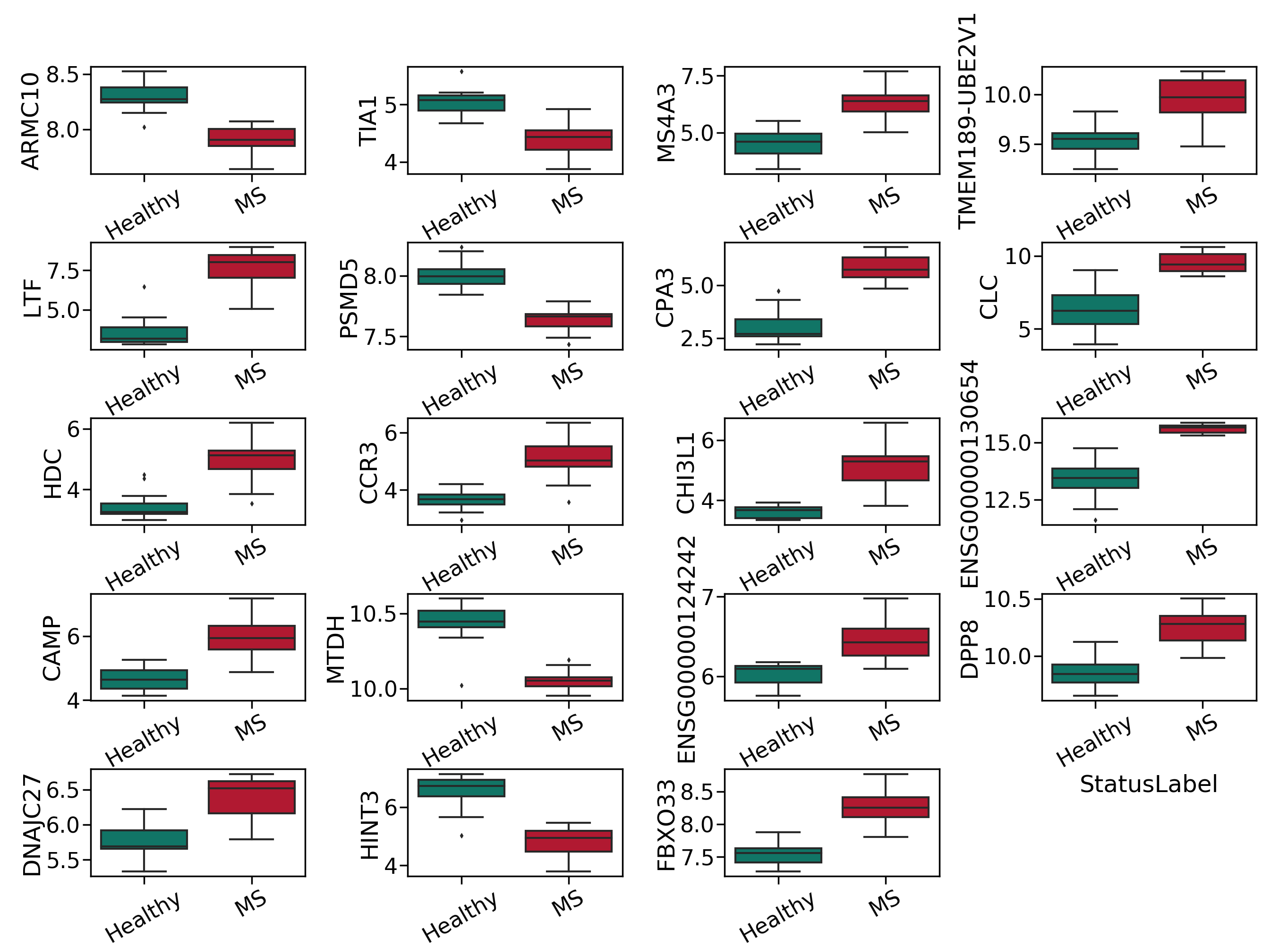}
    \caption{Quantification of the transcripts with most
      significant group differences.}
    \label{fig:gene-trad-trans}
  \end{subfigure}
  \caption{Gene transcription association exploration.}
  \label{fig:gene-trans-exploration}
\end{figure*}

In Figure~\ref{fig:gene-trans-exploration} we study the gene associations
calculated directly from the \ac{EPLS} feature weights. The feature associations
can be calculated from the angle between two transcripts, $\beta$, on the
transcript \ac{EPLS} weight plane. All pairs of $\cos(\beta)$ make up an
association map for the \ac{EPLS} model. The values range from $+1$, for fully
aligned, through $0$, for orthogonal, to $-1$ for antialigned. The \ac{EPLS}
solution constructs a model surface that maximizes the variance between the
$\mat{Y}$ axes. This is the reason why the association map represents an over
accentuated correlation map of the genes depicted.

By selecting a smaller subset of transcripts, here with $q < 0.16$ values, and
calculating the association map we see that the significant transcripts are
roughly divided into two anti-associated clusters. Under the assumption that our
coding genes produce proteins, we may utilize
\texttt{string-db}~\cite{szklarczyk_string_2019} to analyse them. We find that
they enrich for immune system activity in the specific and tertiary granule
lumens. The largest cluster enriches for antifungal humoral response and innate
immune response in mucosa while the second cluster enriches for publications on
long non-coding \ac{RNA} activity. In the second cluster, the gene with the
highest intersection among such publications is the \gene{PSMD5} transcript. The
protein of this gene has also been reported for proteomic quantification of
differentially expressed genes between \texttt{MS} and \texttt{Healthy}
patients~\cite{berge_quantitative_2019}.

\begin{figure*}[t!]
  \centering
  \begin{subfigure}[t]{0.45\textwidth}
    \centering
    \includegraphics[scale=0.3]{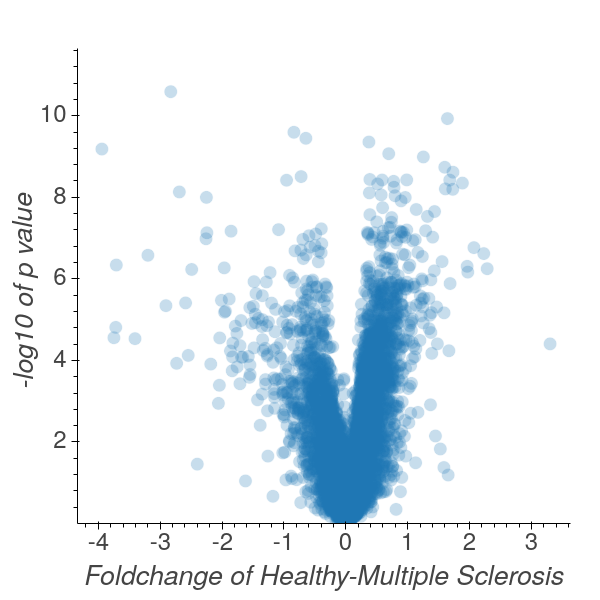}
    \caption{Volcano plot with ttest $p$-values plotted against $log_2$ fold changes.}
    \label{fig:ttest-ms-foldchange-vulcano}
  \end{subfigure}
  ~
  \begin{subfigure}[t]{0.45\textwidth}
    \centering
    \includegraphics[scale=0.3]{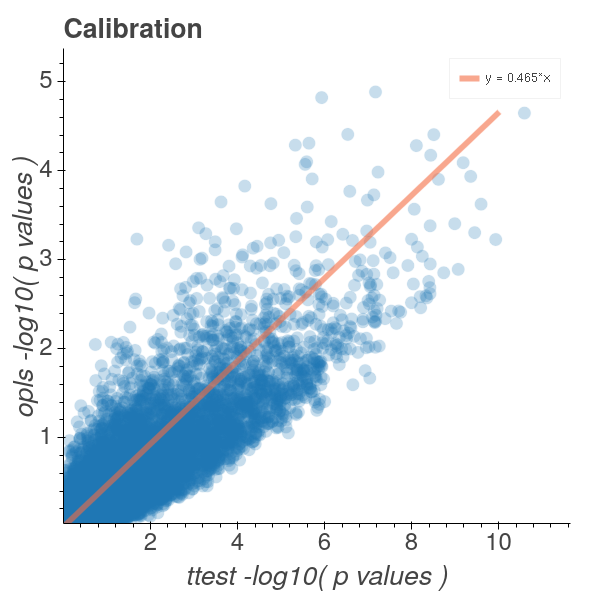}
    \caption{Comparison of the our model $p$-values and traditional t-test derived $p$-values.}
    \label{fig:ttest-ms-foldchange-heteroscedasticity}
  \end{subfigure}
  \caption{MS Comparison of opls and t-test produced $p$-values.}
  \label{fig:ttest-ms-foldchange}
\end{figure*}

From Figure~\ref{fig:ttest-ms-foldchange-vulcano} it is clear that our \ac{EPLS} model
derived $p$-values are conservative estimates when compared to those stemming from a
t-test. From Figure~\ref{fig:ttest-ms-foldchange-heteroscedasticity} it is clear that
the $p$-value distribution is heteroscedastic around the $y \approx 0.5 x$ line in the
logarithmic $p$-value space numerically proving that we are quantifying the same type
of discoveries as the t-test, but with much more conservative estimates of the $p$-values.

\subsection{\ac{TCGA}-\ac{BRCA} and the comparison with a two-way ANOVA} \label{sec:results:tcga-brca}

In this section we study the \ac{TCGA} Breast cancer
dataset~\cite{koboldt_comprehensive_2012} under \ac{EPLS} and a type-2
\ac{ANOVA}. Clinical parameters~\cite{tao_classifying_2019}, as well as
\ac{mRNA} expression data from human female patients. Theses were obtained from
the \ac{TCGA} website~\cite{koboldt_comprehensive_2012}. All of the downloaded
samples originated from Illumina \ac{RNA}-Seq data using the \ac{HTS} 3
\ac{FPKM} workflow~\cite{anders_htseqpython_2015}. In the case where more than
one expression profile belonged to one sample id we randomly selected one of the
profiles. Sample receptor statuses~\cite{ross-innes_differential_2012} were
stored as \gene{ER+}, \gene{ER-}, \gene{PR+}, \gene{PR-}, \gene{HER2+},
\gene{HER2-} or \gene{ND} in a sample journal file. This resulted in $1184$
breast cancer samples~\cite{tjornhammar_ms_2019}.

\begin{figure*}[t!]
  \centering
  \begin{subfigure}[t]{0.45\textwidth}
    \centering
    \includegraphics[scale=0.3]{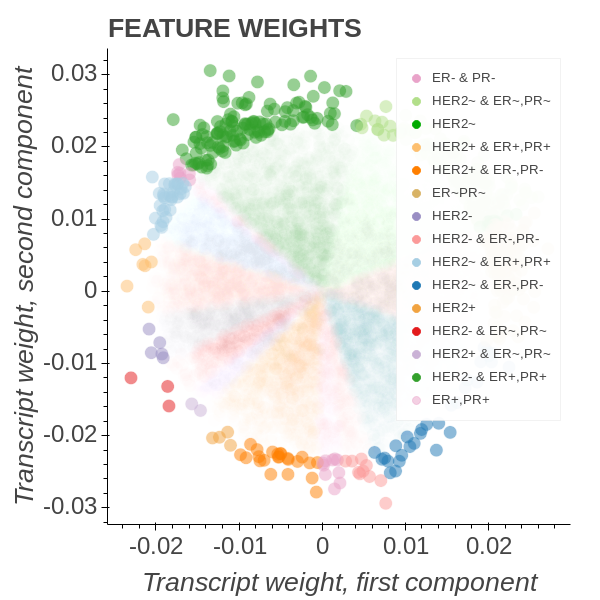}
    \caption{Transcript weight colored by their hormone alignment owners.}
    \label{fig:tcga-brca-trans}
  \end{subfigure}
  ~
  \begin{subfigure}[t]{0.45\textwidth}
    \centering
    \includegraphics[scale=0.3]{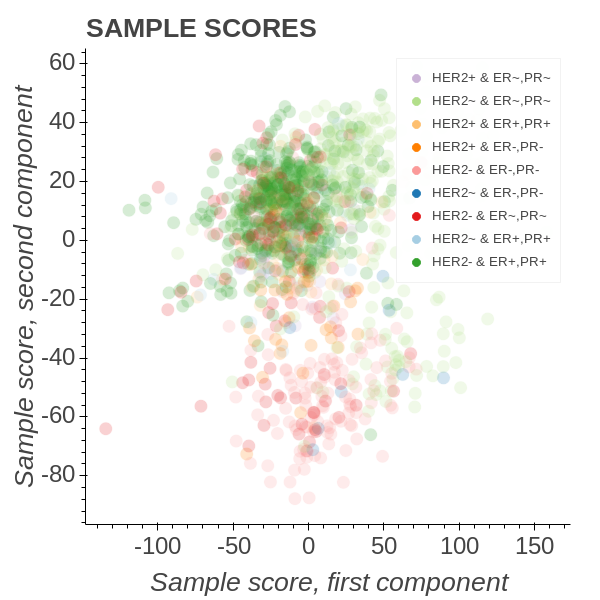}
    \caption{The sample scores show that triple negative patients are
      diametrically opposed to \gene{HER2-} \& \gene{PR+}, \gene{ER+} patients.}
    \label{fig:tcga-brca-sample}
  \end{subfigure}
  \caption{\ac{EPLS} model of the \ac{TCGA} breast cancer data set.}
  \label{fig:tcga-brca-exploration}
\end{figure*}

In Figure~\ref{fig:tcga-brca-exploration} we employed an \ac{EPLS} model of the
form
\[
  \text{Transcript} \sim C(\text{\gene{HER2}}):C(\text{\gene{ERPR}})+C(\text{\gene{HER2}})+C(\text{\gene{ERPR}})
\]
and coloured all the group instances with unique colours. The \gene{HER2} group
describes the \gene{HER2} receptor state while the \ac{ERPR} group describes
whether or not the \gene{ER} and \gene{PR} receptors are in a particular state.
Thus triple-negative patients will belong to the \gene{HER2-} \& \gene{PR-},
\gene{ER-} intergenetion and triple positive to the \gene{HER2+} \& \gene{PR+},
\gene{ER+}. It is well known that triple-negative patients have a poor survival
outcome for breast cancer and are interesting to study in this context.
Figure~\ref{fig:tcga-brca-sample} shows that triple-negative patients amass in
the most southern parts of the lower quadrants. Diametrically opposed we find
the patients bearing the interaction combination belonging to \gene{HER2-} \&
\gene{PR+}, \gene{ER+}. We note that the triple-positive group form an angle of
roughly $110$ degrees to the triple-negative group. From here on the call the
triple-negative towards \gene{HER2-}, \gene{PR+}, \gene{ER+} axis the high-risk
axis.

Among the genes with the strongest association to the high-risk axis, we find
\gene{ESR1}, \gene{CXXC5}, \gene{GREB1}, \gene{FOXA1} and \gene{GATA3}. They
enrich~\cite{szklarczyk_string_2019} for estrogen-dependent gene expression in Reactome as well as
the prostate gland and uterus development in biological processes~\cite{mi_panther_2018}. We
only have women in the data so we safely assume that the prostate gland
enrichment is due to the interaction of these genes with the progesterone
receptor in men.

\begin{figure*}[t!]
  \centering
  \begin{subfigure}[t]{0.45\textwidth}
    \centering
    \includegraphics[scale=0.3]{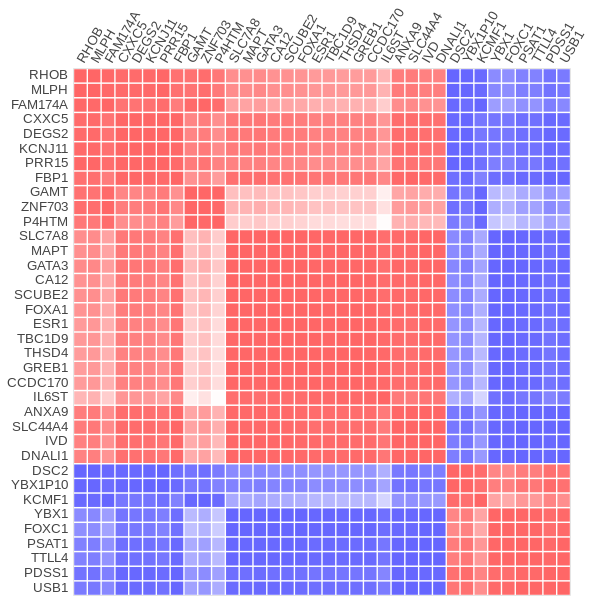}
    \caption{Association map of transcripts describing the variation along the
      triple negative axis}
    \label{fig:tcga-brca-assoc}
  \end{subfigure}
  ~
  \begin{subfigure}[t]{0.45\textwidth}
    \centering
    \includegraphics[scale=0.5]{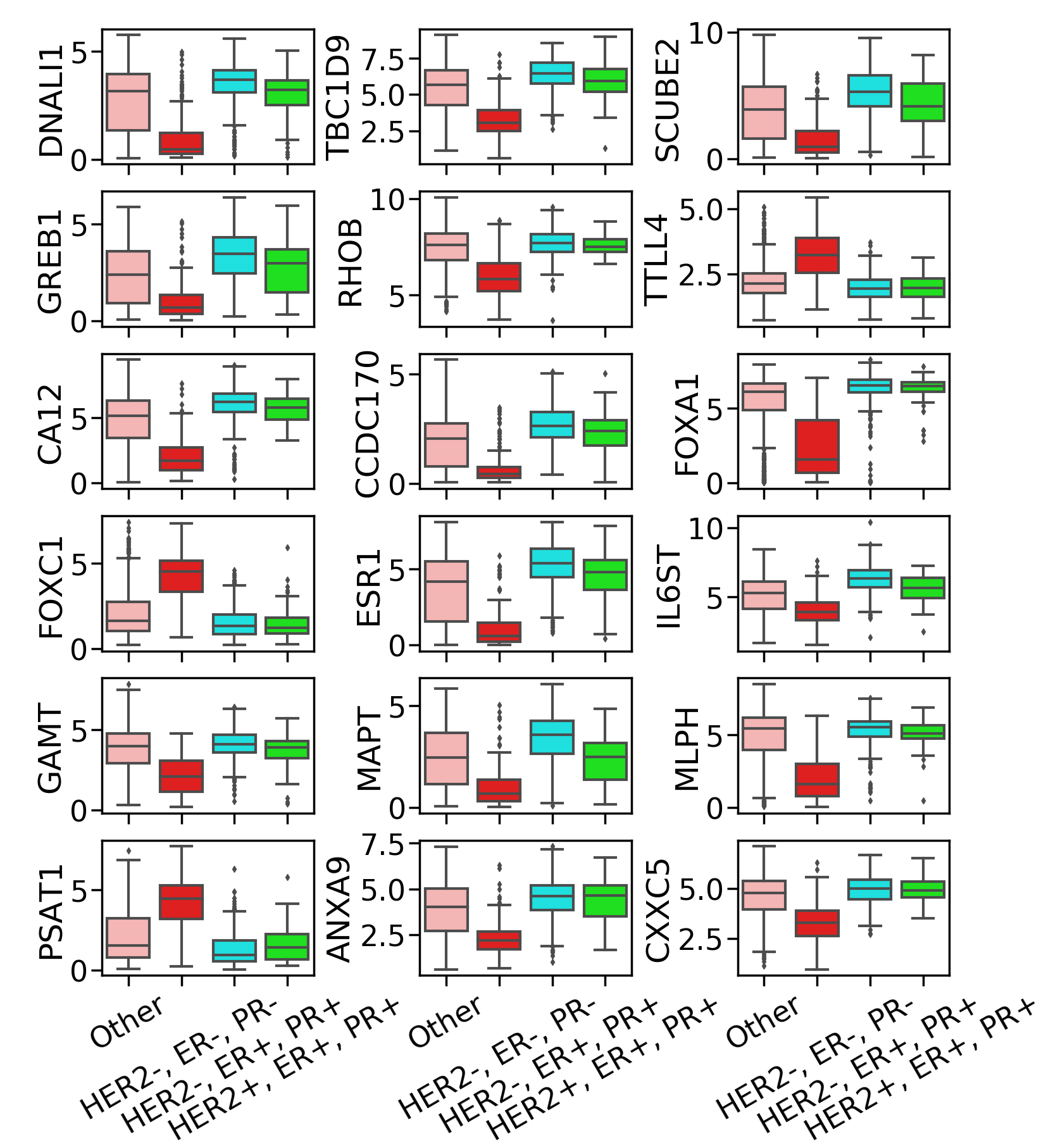}
    \caption{Box plots of most significant transcripts \gene{HER2-}, \gene{PR+},
    \gene{ER+} patients exhibit starkest differences towards triple negatives}
    \label{fig:tcga-brca-boxplot}
  \end{subfigure}
  \caption{\ac{EPLS} model association map of the \ac{TCGA} breast cancer data set.}
  \label{fig:tcga-brca-assoc-exploration}
\end{figure*}

In Figure~\ref{fig:tcga-brca-boxplot}, we find the traditional quantification of the
transcript data. Triple-negative patients are remarkably different from the
other groups. We also show both triple-positive as well as \gene{HER2-},
\gene{PR+} and \gene{ER+} patient groupings. The \ac{EPLS} model in
Figure~\ref{fig:tcga-brca-sample} conveyed the information that \gene{HER2-},
\gene{PR+}, \gene{ER+} combination are anti associated to the triple negatives
and we can also see that the biggest difference in the boxplots of
Figure~\ref{fig:tcga-brca-boxplot} is found between the triple-negative and
\gene{HER2-}, \gene{PR+}, \gene{ER+} group.
It is also becomes apparent that the gene cluster associated with being up in
triple-negative patients belong to \gene{PSAT1}, \gene{TTL4}, \gene{FOXC1},
\gene{USB1}, \gene{PDSS1}, \gene{YBX1}, \gene{KCMF1}, \gene{YBX1P10} and
\gene{DSC2}. It constitutes a novel cluster for describing triple-negative
patients and it is only when evaluated along the high-risk axis that the full
picture emerges.

\begin{figure*}[t!]
  \begin{subfigure}[b]{0.5\linewidth}
    \centering
    \includegraphics[scale=0.3]{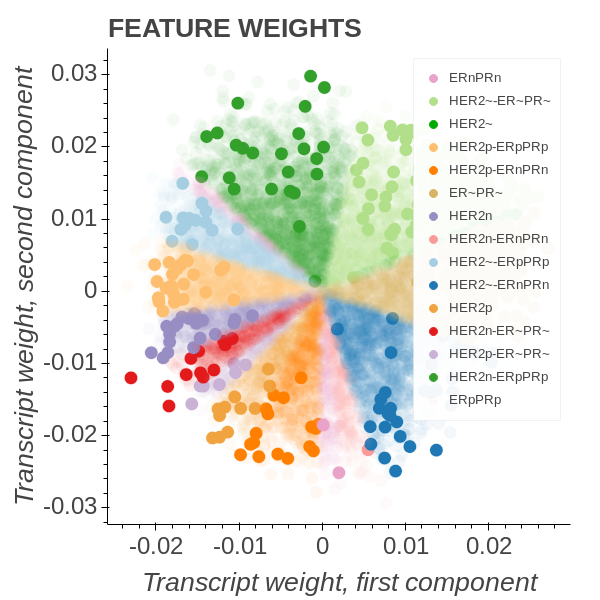}
    \caption{The $p$ values produced by the \ac{EPLS} model as well as the
      intermittent $o$ and final $q$ values.}
    \label{fig:trans-weight-type-2}
  \end{subfigure}
  \begin{subfigure}[b]{0.5\linewidth}
    \begin{subfigure}[b]{0.5\linewidth}
      \centering
      \includegraphics[scale=0.3]{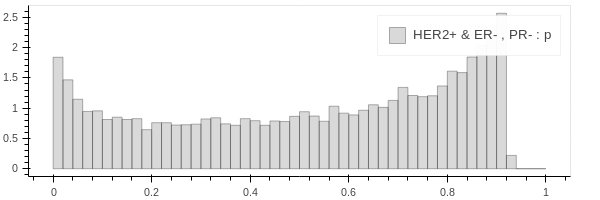}
    \end{subfigure}

    \begin{subfigure}[b]{0.5\linewidth}
      \centering
      \includegraphics[scale=0.3]{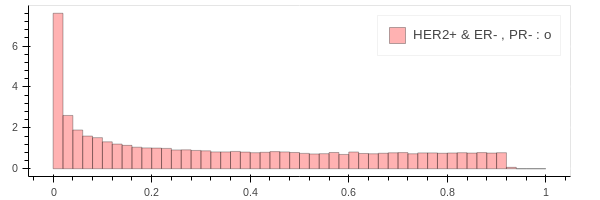}
    \end{subfigure}

    \begin{subfigure}[b]{0.5\linewidth}
      \centering
      \includegraphics[scale=0.3]{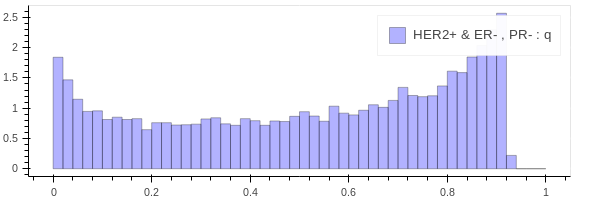}
    \end{subfigure}
    \caption{\ac{EPLS} model with all transcripts having two-way \ac{ANOVA} $q < \frac{0.05}{10455}$ shown with full opacity.}
    \label{fig:trans-type-2-stat-vals}
  \end{subfigure}

\end{figure*}

These results would not be complete without mentioning the statistical properties of the
weights. Since there is no warranty that \ac{EPLS} models converge on any
specific residual distribution this can become convoluted. The maximisation of
variance on the common subspace plane implies that a solution would minimize the
\ac{EPLS} models squared residual. In using our formalism we can study the same
formula using both an \ac{ANOVA} as well as an \ac{EPLS} model. In
Figure~\ref{fig:trans-type-2-stat-vals} we show the same statistical model,
\[
  \text{Transcript} \sim C(\text{\gene{HER2}}):C(\text{\gene{ERPR}})+C(\text{\gene{HER2}})+C(\text{\gene{ERPR}})
\]
produces qualitatively similar results.

In Figure~\ref{fig:trans-weight-type-2} depicts the \ac{EPLS} model where all
the interaction categories of
\[
  \text{Transcript} \sim C(\text{\gene{HER2}}):C(\text{\gene{ERPR}})
\]
are shown. The transcripts
are also modelled using a two-way \ac{ANOVA} and all transcripts having an
interaction with a $q < \frac{0.05}{10455}$ are depicted in full opacity. Two
realizations are immediate. The first is that the \ac{ANOVA} results are
qualitatively similar to those of the \ac{EPLS}, with a large density of
significant \ac{ANOVA} transcripts being found on the rim of the \ac{EPLS}
weight plane. Our \ac{EPLS} model, however, offers an improvement in that we may
determine not just the specific interaction pair associated with which
transcript but also the direction. From our model, we can recognize that
\gene{TTLL4} will be scaling in magnitude with the triple-negative state as well
as being significantly different from the other groupings. The traditional box
plot quantification in Figure~\ref{fig:tcga-brca-boxplot} is important but not
essential. We further explore our \ac{EPLS} $p$-values and it is clear from
Figure~\ref{fig:trans-weight-type-2} that we have a large accumulation of
weights near the centre. This also means that we are more prone to false
discoveries near the centre. One can directly correct for this in the $q$ value
calculation by assuming that the \ac{FDR} scale linearly with the \ac{FR} of the
list of $p$ values, which is quantified by:
\[
  \text{fr}(p) = \frac{\text{ranks}(p)}{\text{length}(p)}
\]
Instead of assuming a constant \ac{FDR}~\cite{storey_statistical_2003} across
$p$ values, often set equal to $1$, for the $q$ value calculation we set it to
scale using the \ac{FR} formula. These scaled $p$ values are here dubbed $o$
values and are only used to show the effect of this scaling. The final $q$
values are showed in Figure~\ref{fig:trans-type-2-stat-vals}. We recognize that
the \ac{EPLS} transcripts with significant interaction all belong to the rim of
the categories defined by the axis studied (here the far edge of the
\gene{HER2+}, \gene{PR-}, \gene{ER-} axis).

\begin{figure*}[t!]
  \centering
  \begin{subfigure}[t]{0.45\textwidth}
    \centering
    \includegraphics[scale=0.3]{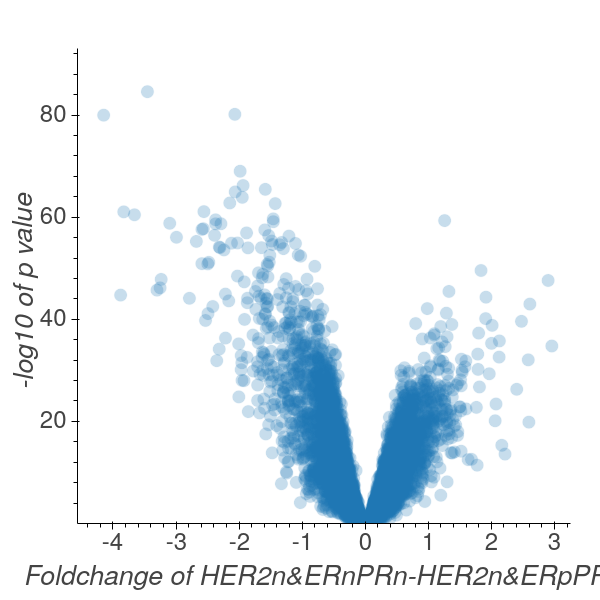}
    \caption{Volcano plot with ttest $p$-values plotted against $log_2$ fold changes.}
    \label{fig:ttest-tcgabrca-foldchange-vulcano}
  \end{subfigure}
  ~
  \begin{subfigure}[t]{0.45\textwidth}
    \centering
    \includegraphics[scale=0.3]{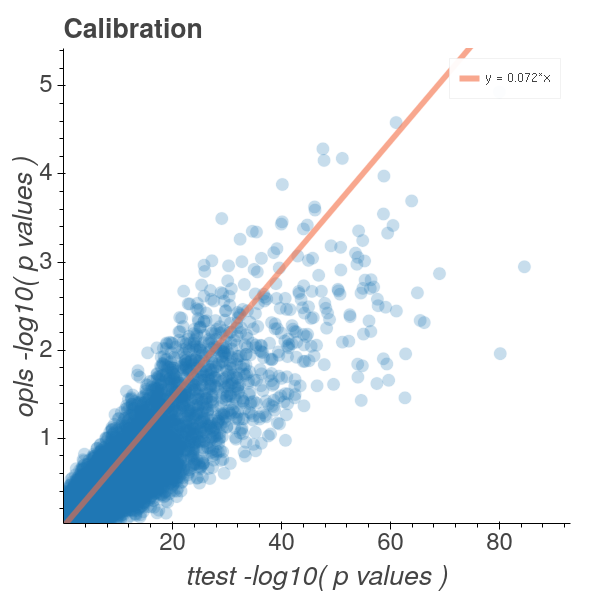}
    \caption{Comparison of the our model $p$-values and traditional t-test derived $p$-values.}
    \label{fig:ttest-tcgabrca-foldchange-heteroscedasticity}
  \end{subfigure}
  \caption{TCGA Comparison of \ac{EPLS} and t-test produced $p$-values.}
  \label{fig:ttest-tcgabrca-foldchange}
\end{figure*}

By inspecting Figure~\ref{fig:ttest-tcgabrca-foldchange} we can see that the
axis found also generates the high significance interaction using a t-test. One
could numerically perfom all pairwise t-tests, we will refrain from doing, in
order to draw a conclusion about wether this axis has the highest t-test
significance. Instead we note that, from Figure~\ref{fig:tcga-brca-trans}, the
axis grouping containing the triple negative samples as well as their
diametrically opposed group is the \gene{HER2-\&ER-},\gene{PR-}
\gene{HER2-\&ER+},\gene{PR+} pair and not the triple positive group. The t-test
significance magnitude drops to half of that obtained from the axis reported as
the most important axis when changed to triple negatives paired against triple
positives. From Figure~\ref{fig:ttest-tcgabrca-foldchange-heteroscedasticity}
the method remains heteroscedastic around a power law relation. However here our
$p$-values roughly fall on the $y \approx x*0.072 $ line in log space. As such we
remain conservative in our $p$-value quantification with respect to a t-test along
a comparable axis. In this case our $p$-values are still of the same order of
magnitude as in Figure~\ref{fig:ttest-ms-foldchange-vulcano}. This is due to the
fact that they are computed from the transcript position and the transcript
density. Since this is decoupled from the samples in the \ac{EPLS} model we can
evaluate it across an arbitrary attribute axis. The transcript density is defined
by roughly the same amount of transcripts in the Multiple Sclerosis and Breast
Cancer cases and the transcript $p$-value magnitudes is therefore comparable and
of the same order of magnitude for both of the \ac{EPLS} models. The \ac{EPLS} weight
alignment become more reliable as more samples are added but this information is
not directly conveyed to the \ac{EPLS} $p$-values. We note that, as in
Figure~\ref{fig:ttest-ms-foldchange-heteroscedasticity}, that our $p$-values are
conservative. By noting that we are projecting onto a spherical solution space
(2-dimensional) where the opls axis (1-dimensional) is set by the sample space
we realize that our axis $p$-values will calibrate towards t-test
derived $p$-values, across the same sample grouping axis as in
equation~\ref{eq:powerlaw}
\begin{equation}
p_{opls,axis} \approx p_{t-test,axis}^{\sqrt{\frac{N_{samples}}{2\pi}}}
\label{eq:powerlaw}
\end{equation}

The main advantage of our method is that we can, using an unbiased method, identify the group
that is diametrically opposed to the disease group.

\subsection{Diabetes Microarray Analysis} \label{sec:results:t2d}

Type 2 Diabetes is a metabolic disease on the rise in the general populous. It is
furthermore influenced by both lifestyle choices as well as genetic factors.
Understanding the detailed balance required for this disease phenotype to emerge
is important to achieve better therapies and quality of life of those
affected by it. The understanding of altered metabolism is also of great
importance since many diseases are associated with altered cell and tissue
metabolism~\cite{mootha_pgc-1-responsive_2003}.

There is an increasing amount of evidence that type 2 diabetes \ac{T2D}, or
\ac{DM2}, disease phenotype constitutes several underlying metabolic disease
phenotypes~\cite{ahlqvist_novel_2018}.

Here we have chosen to study human \ac{DM2} data from the Broad
Institute~\cite{mootha_pgc-1-responsive_2003}. Suspecting that we have more than
a single type of \ac{DM2} patients we first segmented the data. This was done by
employing approximative density clustering as implemented in our
package~\cite{tjornhammar_impetuous-gfa_2019}. Because of the limited number of
samples $(18)$ we stopped once we had obtained the two most separated sample
groups. We thereby obtain two well-separated \gene{DM2} cluster types that we
name \gene{C1.DM2} and \gene{C2.DM2} where the cluster index is the inverted
cluster size rank.

\begin{figure*}[t!]
  \centering
  \begin{subfigure}[t]{0.45\textwidth}
    \centering
    \includegraphics[scale=0.3]{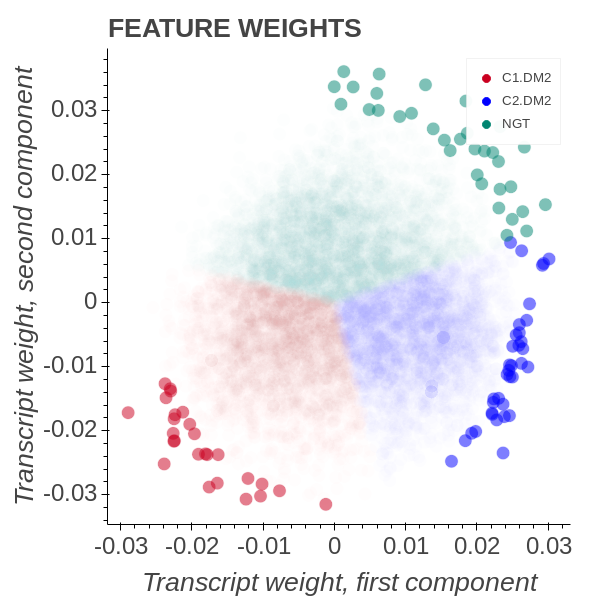}
    \label{fig:ngt-trans-weights}
  \end{subfigure}
  ~
  \begin{subfigure}[t]{0.45\textwidth}
    \centering
    \includegraphics[scale=0.3]{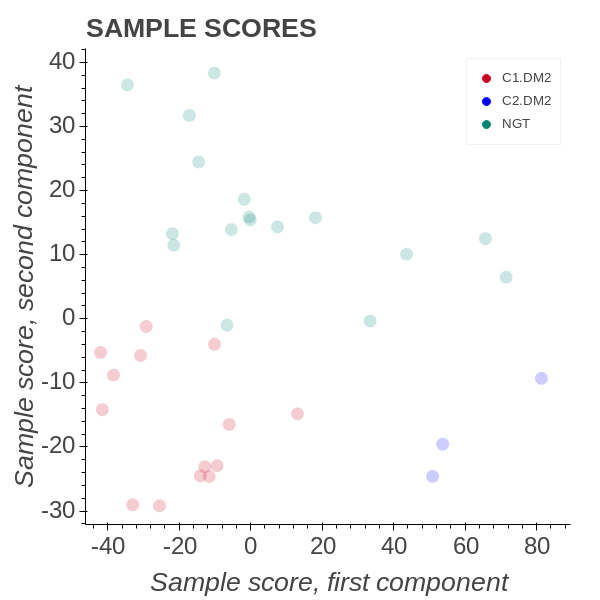}
    \label{fig:ngt-trans-samples}
  \end{subfigure}
  \caption{\ac{T2D} \& \ac{NGT} \ac{EPLS} model with all transcripts.}
  \label{fig:ngt-trans-exploration}
\end{figure*}

\begin{figure*}[t!]
  \centering
  \begin{subfigure}[t]{0.45\textwidth}
    \centering
    \includegraphics[scale=0.3]{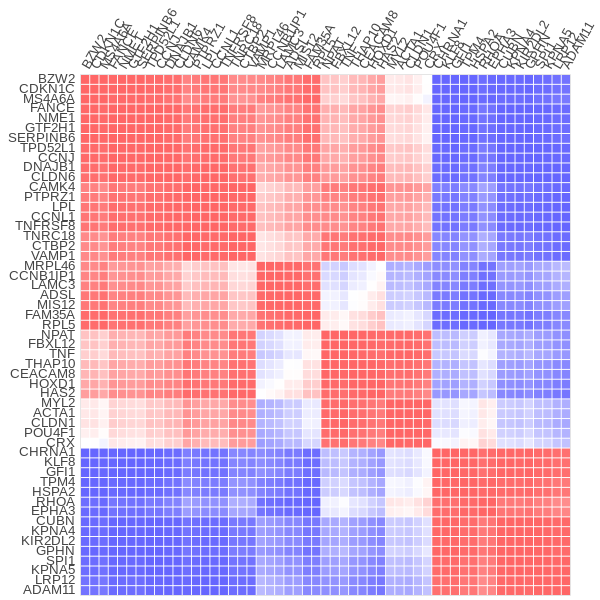}
    \caption{All the edge weight associations below $q < 0.005$}
    \label{fig:low-q-assoc-map}
  \end{subfigure}
  ~
  \begin{subfigure}[t]{0.45\textwidth}
    \centering
    \includegraphics[scale=0.3]{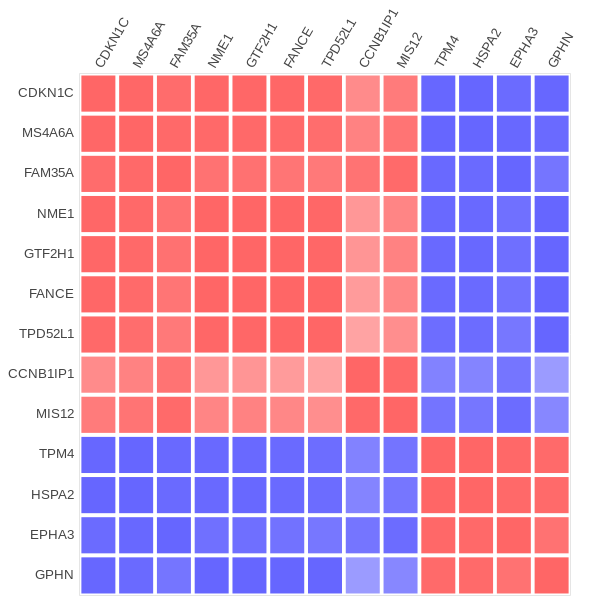}
    \caption{The association map of the largest \ac{DM2} group and \ac{NGT}
      variation, $q < 0.0001$}
    \label{fig:ngt-trans-samples}
  \end{subfigure}
  \caption{Complete \ac{EPLS} association maps.}
  \label{fig:dm2-ngt-variation}
\end{figure*}

In Figure~\ref{fig:ngt-trans-exploration} we see that the second cluster
consists of only three samples. They form a group well-separated from the rest
but lying close to orthogonal to the \ac{NGT} group.  The first cluster forms an
axis with the \ac{NGT} group with weights being positioned opposed to one
another.
Using our \ac{EPLS} model we would believe that they constitute a
smaller \ac{DM2} phenotype subset with distinct metabolic alterations. By
studying the $p$-values associations on the rim of the entire \ac{EPLS} graph we
find that \gene{FAM35A} have $q < \frac{0.05}{N_{transcripts}}$ and aligning
with the \gene{C1.DM2} axis. \gene{FAM35A} is known to be a favourable
prognostic marker of glioma and colo-rectal
cancer~\cite{uhlen_tissue-based_2015}. By allowing ourself to study genes with
$q < 0.005$ we obtain the association map in Figure~\ref{fig:low-q-assoc-map}.

In Figure~\ref{fig:dm2-ngt-variation} we can clearly discern four transcript
clusters. The first two correspond to transcripts that are positively
associating with being healthy. Those transcripts enrich for Cyclin C activity,
which is known for controlling nuclear cell division~\cite{galderisi_cell_2003}.
The third cluster is positioned near the centre of the association map on the
left and is close to orthogonal to the other clusters.

This cluster consists of transcripts that enrich~\cite{szklarczyk_string_2019}
for striated muscle contraction~\cite{jassal_reactome_2019} and organ
morphogenesis~\cite{mi_panther_2018}. Most of the enrichment signal is, however,
coming from the \gene{MYL2}, \gene{ACTA1}, \gene{CLDN1}, \gene{POU4F1},
\gene{CRX} and \gene{TNF} group. This third cluster belongs to the \gene{C2.DM2}
group indicating that these three individuals, albeit being diabetic, might have
dramatically different muscle metabolism.

The first and biggest cluster belong is the \gene{C1.DM2} samples visible in
Figure~\ref{fig:dm2-ngt-variation}. Four transcripts reach high significance
and together enrich for \gene{ATP} binding activity. The $9$ high significance
transcripts aligning with this axis enrich~\cite{szklarczyk_string_2019} for
\ac{DNA} repair and \ac{DNA} metabolism. Taken together with the full
\gene{C1.DM2-NGT} axis transcripts enrich for activity in the condensed nuclear
chromosome. By allowing ourselves to be less strict and studying enrichment at
lower confidence corresponding to the \gene{C1.DM2-NGT} transcripts in
Figure~\ref{fig:low-q-assoc-map}.

Approximate density clustering is similar to agglomerative hierarchical
clustering but instead of building a distance hierarchy we try to find the best
sample grouping describing the sample space.

\begin{figure*}[t!]
  \centering
  \begin{subfigure}[t]{0.45\textwidth}
    \centering
    \includegraphics[scale=0.3]{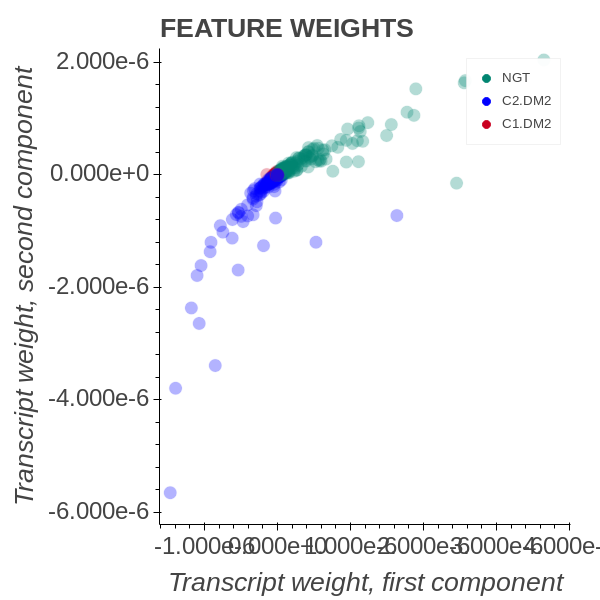}
    \label{fig:ngt-trans-pca-weights}
  \end{subfigure}
  ~
  \begin{subfigure}[t]{0.45\textwidth}
    \centering
    \includegraphics[scale=0.3]{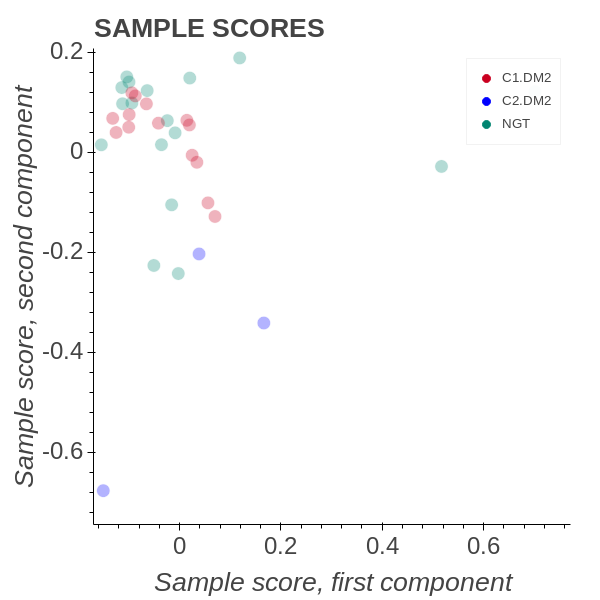}
    \label{fig:ngt-trans-pca-samples}
  \end{subfigure}
  \caption{\ac{T2D} \& \ac{NGT} \ac{PCA} model with all transcripts.}
  \label{fig:ngt-trans-pca-exploration}
\end{figure*}

All other things being equal we can compare
Figure~\ref{fig:ngt-trans-exploration} and
Figure~\ref{fig:ngt-trans-pca-exploration} and see that \ac{EPLS}, by its very
nature, better separates the data. In Figure~\ref{fig:ngt-trans-pca-exploration}
we see that the three samples who have their own cluster in
Figure~\ref{fig:ngt-trans-exploration} also correspond to outliers when
employing \ac{PCA}.

\section{Discussion} \label{sec:discuss}

It should be noted that the method is meant to be used as an exploratory
descriptive method. There has been no attempt to validate its efficacy for
inference. KNN is also a method that can be implemented in multiple ways and in
this specific case 1-NN was employed simply because the PLS operations are taken
beforehand, ostensibly removing noise from each ``centroid''.

An alternate way of studying $Y$ space variance would be via discriminant
analysis, as suggested by research related to~\cite{trygg_orthogonal_2002}
and~\cite{wold_partial_2006}. \ac{OPLS}-DA solves this but cannot relate a
specific feature to a specific attribute.

Leaving the general discussion aside we also have comments regarding the use
case specific results in the following three subsections. These discussions are
to elaborate on the suitability of the method uncovered results.

\subsection{Multiple Sclerosis} \label{sec:discuss:ms}

Regarding the multiple sclerosis use case we see from
Figure~\ref{fig:opls-hms-exploration} in Section~\ref{sec:results:ms} that the
major pathways deduced are complex I biogenesis and neutrophil granulation,
pathways which are affiliated with increased metabolism and immune system
activity. Compartment enrichment analysis, also utilizing a two-sided Fisher
exact test, of our $p$-values tells us that the dataset is mainly active in the
cytosol, nucleoplasm and ficolin-1-rich granule membrane. This conveys the
picture that we are dealing with neutrophil degranulation and metabolism of
surface membranes. This is aligned with the common view that \ac{MS} is
associated with demyelination of axon myelin. Roughly translating into altered
lipid metabolism by the immune system.

Our modelling scheme shows weaker statistical power than other methods but
recovers fundamental and broad descriptive knowledge about the dataset. We also
identify a small subset of significant genes that are differentially expressed
in the cohort and share an intersection with proteomic analysis. Both the
proteomic derived set as well as our \ac{EPLS} transcriptome analysis does not
share a large common intersecting set with the differentially expressed genes
published together with the \ac{GWAS} dataset that we employed.
This analysis method should as such be viewed as a complementary method for
producing a data-driven story of the cohort as well as
targets of interest for use in further biological validation. In this section,
we have shown how the $\mat{Y}$ matrix alignments translate into the feature
space of \ac{EPLS} models.

Looking at Figure~\ref{fig:gene-cosine-assoc} and allowing ourself down into the
first $100$ significant \ac{EPLS} genes, with a $p < 0.0002$ value, we see a
common intersection set between our results and
the~\cite{berge_quantitative_2019} proteins. These are \gene{TES}, \gene{CLTC},
\gene{SIN3A} as well as \gene{PSMD5} and known to influence \gene{RUNX1}
transcription pathways.

\subsection{Breast Cancer} \label{sec:discuss:tcga-brca}

In the breast cancer use case; the \ac{FOX} \gene{C1} transcript is thought to
be a possible target in nasopharyngeal carcinoma~\cite{ou-yang_forkhead_2015}.
\gene{FOXA1} is similarly known to be correlated with \gene{GATA3}, Progesteron
receptor protein expression and clinical outcomes in breast
cancer~\cite{ross-innes_differential_2012} as well as prostate
cancer~\cite{barbieri_exome_2012}. However, some of the associations in
Figure~\ref{fig:tcga-brca-assoc} are new and might be interesting targets for
further biological validation. It is worth further emphasis that these gene
transcripts correspond to the outliers lying on the axis connecting the
triple-negative and the \gene{HER2-}, \gene{PR+}, \gene{ER+} in the \ac{EPLS}
feature weight plane.

\begin{figure*}[t!]
  \begin{center}
  \includegraphics[scale=0.4]{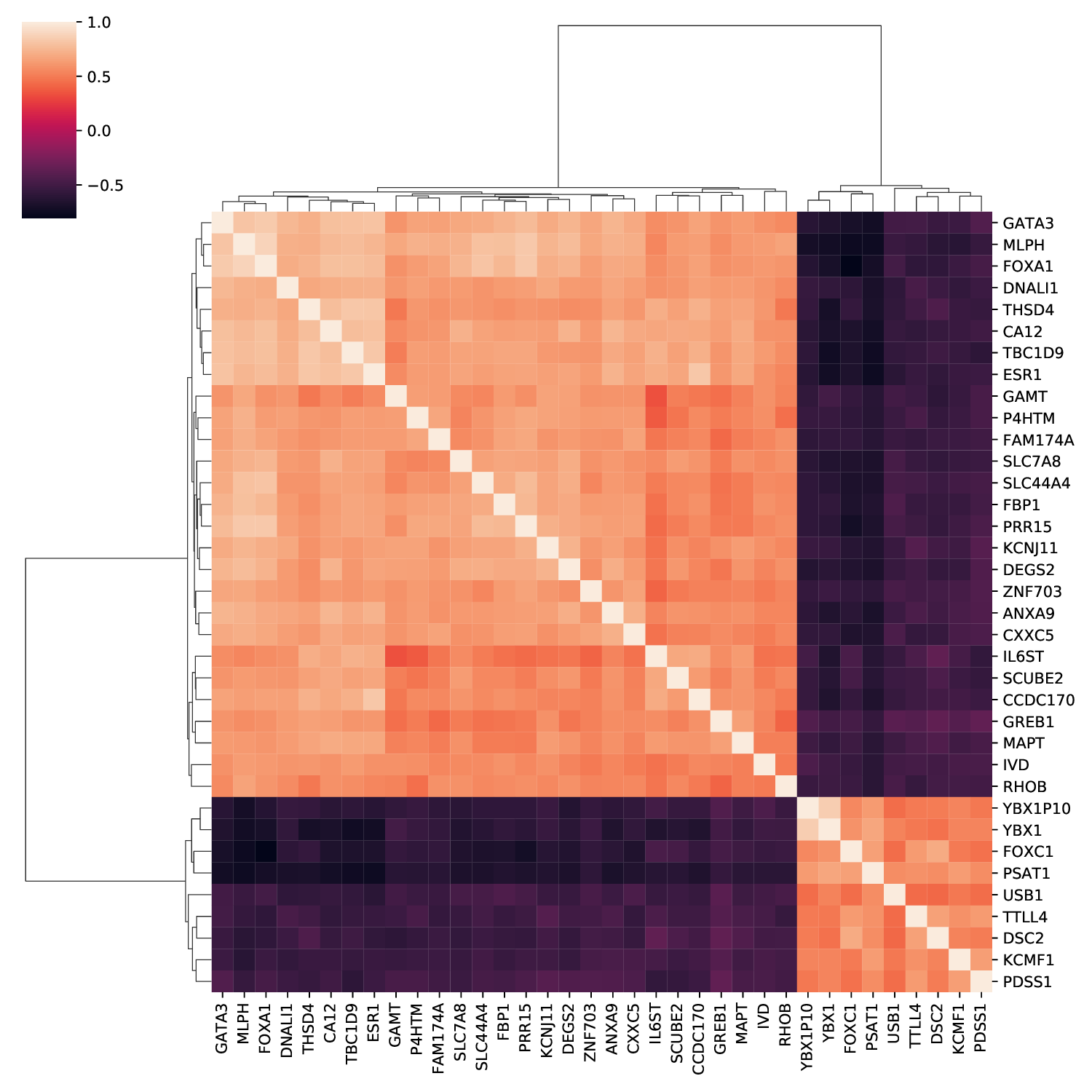}
  \caption{Clustered correlation heatmap of the genes identified by the EPLS method for the triple-negative axis.}
  \label{fig:tcga-brca-clustermap}
  \end{center}
\end{figure*}

One can compare the association map in
Figure~\ref{fig:tcga-brca-assoc-exploration} to an alternate method, such as a
clustered correlation heatmap. In Figure~\ref{fig:tcga-brca-clustermap} we can
see a clustermap providing the qualitative same results, while residing on a
fundamentally different mathematical approach. Meaning that obtaining a gene
cosine angle value close to 1 will obtain a rank correlation close to 1 and
that, similarly, anti-aligned genes become anti-correlated.

\subsection{Type 2 Diabetes} \label{sec:discuss:t2d}

Hierarchical modelling or agglomerative hierarchical clustering are well-known
methods of partitioning one's data. Traditionally, building hierarchical models
are done via curating knowledge and manually constructing a
hierarchy~\cite{jassal_reactome_2019,ahlqvist_novel_2018}. One can also choose
to computationally construct hierarchical data structures using agglomerative
hierarchical clustering~\cite{mehta_high-bias_2019} or simply try to find the
grouping that will separate the density the
most~\cite{tjornhammar_impetuous-gfa_2019}.

Guided by Figure~\ref{fig:dm2-ngt-variation} we see an
enrichment~\cite{szklarczyk_string_2019} for \ac{GTP} metabolism in the
mitochondria. We know that mitochondrial \ac{GTP} metabolism is essential for
glucose-dependent insulin secretion~\cite{kibbey_mitochondrial_2007}. It is
known that this will lead to disrupted oxidative phosphorylation and an altered
mitochondrial membrane potential~\cite{mehta_high-bias_2019}. Furthermore,
mitochondria maintain oxidative phosphorylation by creating a membrane
potential, also implying an altered membrane potential will lead to altered
oxidative phosphorylation~\cite{momcilovic_vivo_2019}. This together is
indicative that \ac{ATP} synthesis might be inhibited, for the diabetics, in the
mitochondria~\cite{sivitz_mitochondrial_2010}. These findings are qualitatively
similar to those reported by~\cite{mootha_pgc-1-responsive_2003}. This is indeed
good news since we are studying people whom, by definition, have inhibited
metabolic regulatory control of glucose and insulin.

\section{Conclusions} \label{sec:conclusions}

We have developed a new analysis method and code library, as
described in Section~\ref{sec:method} and~\ref{sec:impl}, for use in
bioinformatic analysis. We have demonstrated the use of \ac{EPLS} models for
transcriptomic analysis on three different \ac{mRNA} data sets, \ac{MS}
(Section~\ref{sec:results:ms}), \ac{TCGA}-\ac{BRCA}
(Section~\ref{sec:results:tcga-brca}) and \ac{T2D}
(Section~\ref{sec:results:t2d}).

Our models produce unbiased sensible descriptions of the endpoint endogen states
where we can relate what we see in the data to known properties of the human
patient samples that we studied, as discussed in Section~\ref{sec:discuss}. We
can see that \ac{PLS}, through clever encoding and alignment produces sample attribute
specific insights. We have shown how the alignment of the sample to feature scores
translate into the ability to interpret feature weights in the context of the
sample weights. As such the realization that the weights of the feature and
sample descriptor matrices are aligned facilitates the use of the feature
weights directly in traditional pathway analysis methods.  This also means that
a high weight in the direction of a predictor axis will equate to large,
between-group, difference quantifications lying on that axis. In contrast to,
e.g. DSeq2 we can compare more than two known covariates against each other in a
many to many comparison.

From this, several new and sensible interpretations of the data emerge. We can
also confirm previously known properties of the samples. We find a gene cluster
which follows the triple-negative breast cancer patients which have interesting
suggestion targets for continued validation.

\ac{OPLS} models are employed in such diverse research fields as financial
forecasting~\cite{huang_integrating_2010}, brain
imaging~\cite{abdi_partial_2010} as well as environmental spatial
characterization and forecasting~\cite{thioulouse_multivariate_1995}.

We believe
, as such, that this method is suitable as decision support for providing global
suggestions of future study in high entropy domains.

To our knowledge, looking
at other \ac{OPLS} model applications,  no one explores the orientation, or
vector alignment, properties of such models.

\newpage

\bibliographystyle{RS}
\bibliography{epls}

\end{document}